\documentclass[namedreferences]{solarphysics}

\usepackage[hyperref,optionalrh]{spr-sola-addons} 
\usepackage{graphicx}        
\usepackage{color}           
\usepackage{breakurl}        

\newcommand{\grl}{    {\it Geophys. Res. Lett.}}

\newcommand{\jgr}{    {\it J. Geophys. Res.}}

\newcommand{\solphys}{{\it Solar Phys.}}

\chardef\us=`\_

\begin{document}

\begin{article}
\begin{opening}

\title{Dependence of Intensities of Major Geomagnetic Storms (Dst$\le$-100 nT) on Associated Solar Wind Parameters}

\author[addressref={aff1,aff2},corref,email={Legm@cma.gov.cn}]{\inits{Le}\fnm{Le}~\lnm{Gui-Ming}}
\author[addressref={aff2}]{\fnm{Liu}~\lnm{Gui-Ang}}
\author[addressref={aff1}]{\fnm{Zhao}~\lnm{Ming-Xian}}

\address[id=aff1]{Key Laboratory of Space Weather, National Center for Space Weather, China Meteorological Administration, Beijing, 100081, P.R. China}
\address[id=aff2]{School of Physics Science and Technology, Lingnan Normal University, Zhanjiang, 524048, P.R. China}

\runningauthor{G.-M. Le et al.}
\runningtitle{Dependence of Geomagnetic storms on Solar Wind Parameters}

\begin{abstract}
A geomagnetic storm is the result of  sustained interaction between solar wind with a southward magnetic field and the magnetosphere. To investigate the influence of various solar wind parameters on the intensity of major geomagnetic storm, 67 major geomagnetic storms that occurred between 1998 and 2006  were used to calculate the correlation coefficients (CCs) between the intensities of major geomagnetic storms and the time integrals of southward interplanetary magnetic field $B_s$, solar wind electric field ($E_y$) and injection function (Q) during the main phase of the associated geomagnetic storms. SYM-H$_{min}$ was used to indicate the intensity of the associated major geomagnetic storm, while I($B_z$), I($E_y$) and I(Q) were used to indicate the time integrals of $B_z$, $E_y$ and Q during the main phase of associated major geomagnetic storm respectively. The derived CC between I($B_z$) and SYM-H$_{min}$ is 0.33, while the CC between I($E_y$) and SYM-H$_{min}$ is 0.57 and the CC between I(Q) and SYM-H$_{min}$ is 0.86. The results provide statistical evidence that solar wind dynamic pressure or solar wind density plays a significant role in transferring solar wind energy into the magnetosphere, in addition to the southward magnetic field and solar wind speed. Solar wind that has a strong geoeffectiveness requires solar wind dynamic pressure $>$3 nPa or solar wind density $>3$ nPa$/V_{sw}^2$. Large and long duration $B_s$ alone cannot ensure a major geomagnetic storm, especially if the solar wind dynamic pressure is very low, as large and long duration Bs is not a full condition, only a necessary condition to trigger a major geomagnetic storm.
\end{abstract}
\end{opening}

\section{Introduction}
     \label{S-Intro}

A geomagnetic storm is a significant disturbance in the Earth¡¯s magnetic field (e.g. Gonzalez et al., 1994) due to the continuous interaction between solar wind with a southward magnetic field and the magnetosphere.  It is generally accepted that a large and long duration southward interplanetary magnetic field ($B_s$$>$10 nT for more than 3 h) will lead to a major geomagnetic storm (Gonzalez and Tsurutani, 1987). To investigate the effects of various solar wind parameters on the associated geomagnetic storms, many researchers calculated the CCs between the peak values of various solar wind parameters and the intensities of associated geomagnetic storms (e.g. Choi et al 2009, Kane 2005, 2010, Echer et al., 2008a, Ji, et al. 2010, Richardson and Cane 2011, Wu and Lepping 2002, Wu et al. 2016). Of the various solar wind parameters, peak values of $B_s$ and $E_y$ usually have good correlation with the intensity of the associated geomagnetic storm, while solar wind density alone usually has a poor correlation with the intensity of the associated geomagnetic storm. The CC between the intensity of a super geomagnetic storm and peak $B_s$ and $E_y$ has been calculated according to solar wind data and 11 super geomagnetic storms that occurred during solar cycle 23 (Echer et al. 2008b). The CC between peak $B_s$ and the intensity of the super geomagnetic storm was found to be 0.23, and the CC between peak $E_y$ and the intensity of the super geomagnetic storm was also 0.23. However, the CC between peak $B_s$ and the intensity of the super geomagnetic storm in the paper by Meng et al. (2019) was 0.93. Meng et al. (2019) suggested that the high CC between peak $B_s$ and the intensity of the super geomagnetic storm implied that the peak magnitude of Bs determines the strength of a superstorm to a large extent. However, the contribution made by solar wind speed and density to the intensity of a super geomagnetic storm has not been investigated.

A geomagnetic storm comprises a main phase and a recovery phase. The development of a geomagnetic storm depends on the injection term and the decay term of the ring current. The injection term is larger than decay term during the main phase of a geomagnetic storm. Some researchers proposed that the injection term is only a linear function of solar wind electric field with solar wind density playing no role (Burton et al. 1975, Fenrich and Luhmann 1998, O'Brien and McPherron 2000b). Statistical results reported by O¡¯Brien and McPherron (2000) suggested that the solar wind density does not independently drive the ring current. According to solar wind data and the minimum of Dst indices of 11 super geomagnetic storms (Dst$\le$-250 nT) that occurred during solar cycle 23, the CC between I($E_y$) and the minimum of Dst is 0.62 (Echer et al. 2008b). We note that the effect of solar wind density on the intensity of a super geomagnetic storm was not investigated in the paper by Echer et al. (2008b).

Wang et al. (2003) found that the injection term of the ring current, Q, not only depends on the solar wind electric field, but also depends on the solar wind dynamic pressure. Because solar wind dynamic pressure is a function of solar wind speed and density, hence solar wind density is also an important factor for the associated geomagnetic storm. Case studies (Kataoka et al. 2005, Le et al. 2020), global MHD simulations (Lopez et al. 2004), and an impulse response function model (Weigel 2010) suggest that solar wind density is an important parameter modulating the transfer of solar wind energy to the magnetosphere during the main phase of a storm.

There are two different views with regard to the injection term of the ring current of a geomagnetic storm at present. One of the perspectives is that the injection term of the ring current is a linear function of the solar wind electric field. Alternatively, the injection term of the ring current may not only be a function of the solar wind electric field, but also a function of the solar wind dynamic pressure. Now the question is which one is better? To shed light on this discrepancy, the CC between I(Q) and the intensity of the associated major geomagnetic storm should be calculated and compared with the CC between I($E_y$) and the intensity of the associated major geomagnetic storm; this is the objective of the present study.

It is generally accepted that large and long duration southward IMF (Bs$>$10 nT for more than 3 h) will lead to a major geomagnetic storm (Gonzalez and Tsurutani, 1987). Can large and long duration southward IMF alone produce a major geomagnetic storm? To answer this question, the CC between I($B_z$) and the intensity of the associated major geomagnetic storm is calculated in this work. The organization of this paper is as follows. Data analysis is presented in Section 2. The discussion and conclusions are presented in the final section.

The time resolution of the Dst index is 1 hour, while the time resolution of the SYM-H index is 1 minute. The study conducted by Wanliss and Showalter (2006) suggests that the SYM-H index can be used as a high time resolution Dst index. The start and end time of the main phase of a geomagnetic storm can be precisely determined according to the SYM-H index, and then the period of solar wind responsible for the main phase of the associated storm can be precisely determined. Time integrals of various solar wind parameters responsible for the main phases of associated geomagnetic storms can be calculated precisely, and then the CC between the time integrals of various solar wind parameters during the main phases of associated major geomagnetic storms and the intensities of associated major geomagnetic storms can be calculated.

\section{Data Analysis}
\label{S-Data}

\subsection{Solar Wind Data and Geomagnetic Storm Data}

The SYM-H index was obtained from the World Data Center for Geomagnetism, Kyoto (available at http://wdc.kugi.kyoto-u.ac.jp/aeasy/index.html). In this study, 64 solar wind data observed by the ACE spacecraft from 1998 to 2006 (available at ftp://mussel.srl.caltech.edu/pub/ace/level2/magswe/) were used. Solar wind data with a time resolution of 64 s has only been available since 1998; hence, major geomagnetic storms during the period from 1998 to 2006 were used in the present study. Because of the solar wind data gap for some major geomagnetic storms, only the main phases of 68 major geomagnetic storms have solar wind data. As such, 67 major geomagnetic storms were used to study the effects of solar wind parameters on the intensity of major geomagnetic storms in the present study.

\subsection{Time Integrals of Solar Wind Parameters}

We use I($B_z$) to indicate the time integral of $B_z$ during the main phase of a storm, which is calculated as follows:
\begin{equation}\label{eq-01}
I(B_z) = \int_{t_s}^{t_e}B_z \mathrm{d}t
\end{equation}
If $B_z$ is northward, then $B_z$ is set as zero in the calculation of $I(B_z)$. $t_s$ and $t_e$ indicate the start and the end time of associated storm main phase respectively.

Solar wind electric field is calculated by $E_y$=$V_{sw}B_z$, where $V_{sw}$ is the solar wind speed, and $B_z$ is the z-component of IMF. We use I($E_y$) to indicate the time integral of $E_y$ during the main phase of a storm, which is calculated as follows:
\begin{equation}\label{eq-02}
I(E_y) = \int_{t_s}^{t_e}E_y \mathrm{d}t = \int_{t_s}^{t_e}V_{sw}B_z \mathrm{d}t
\end{equation}
If $B_z$ is northward, then $B_z$ is set as zero in the calculation of $I(E_y)$.

Burton et al. (1975) proposed a linear function of the dawn-to-dusk component of solar wind electric field to describe the changes of the pressure-corrected Dst index caused by the energy injection from the solar wind into the ring current as well as the ring current decay,
\begin{equation}\label{eq-03}
  \mathrm{d}Dst^*/\mathrm{d}t = Q(t)-Dst^*/\tau
\end{equation}
where Dst$^*$ is the pressure-corrected Dst index and the contribution made by the magnetopause current has been subtracted in formula (\ref{eq-03}). We directly use $SYM-H$ to substitute Dst in formula (\ref{eq-03}) in the present study. $\tau$ and Q are the decay time and the injection term of the ring current, respectively. Q has the following form,
\begin{equation}\label{eq-04}
  Q = \left\{ \begin{array}{ll}
                0 & B_z \ge 0 \\
                \left|VB_z\right| &  B_z < 0
              \end{array} \right.
\end{equation}
Q proposed by Fenrich and Luhmann (1998) is also a linear function of solar wind electric field with some difference from the Q described in the paper by Burton et al. (1975).

The injection term of the ring current described by Wang et al. (2003) is calculated as follows:
\begin{equation}\label{eq-05}
  Q = \left\{ \begin{array}{ll}
                0 & VB_s \le 0.49 mV/m \\
                -4.4(VB_s-0.49)(P_{sw}/3)^{0.5} & VB_s > 0.49 mV/m
              \end{array} \right.
\end{equation}
Where $P_{sw}$ is solar wind dynamic pressure. It is evident that Q not only depends on solar wind electric field, but also depends on solar wind density.  $I(Q)$, which indicates the time integral of Q during the main phase of a storm, is calculated by the formula listed below.
\begin{equation}\label{eq-06}
  I(Q) = \int_{t_s}^{t_e}Q\mathrm{d}t
\end{equation}.

\section{Results} 
      \label{S-Result}
\begin{figure}
  \centering
  \includegraphics[width=0.8\textwidth]{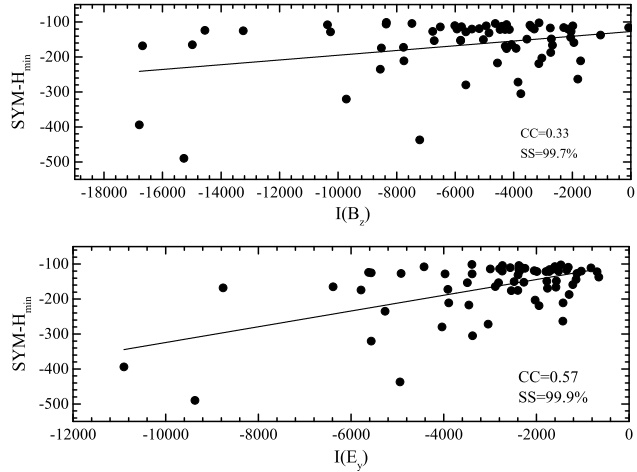}
  \caption{The CC between SYM-H$_{min}$ and $I(B_z)$ (upper panel) and the CC between SYM-H$_{min}$ and $I(E)$ (lower panel)}\label{fig-01}
\end{figure}

For convenience of description, we use CC(X, Y) to indicate the CC betweentwo parameters X and Y. The CC between the minimum of SYM-H and I($B_z$) is calculated and shown in the upper panel of Figure \ref{fig-01}. As shown in Figure \ref{fig-01}, CC(SYM-H$_{min}$, $I(B_z)$) is only 0.33, suggesting that SYM-H$_{min}$ index has a poor correlation with $I(B_z)$, meaning that southward IMF alone makes a small contribution to the intensity of an associated geomagnetic storm. As shown in the lower panel of Figure 1, CC(SYM-H$_{min}$, $I(E_y)$) is 0.57, which is much larger than CC(SYM-H$_{min}$, $I(B_z)$). Because $E_y$=$V_{sw}B_z$, CC(SYM-H$_{min}$, $I(E_y)$) is much larger than CC(SYM-H$_{min}$, $I(B_z)$); thus, solar wind speed is an important parameter for the intensity of an associated major geomagnetic storm. The derived statistical significance (ss) of CC(SYM-H$_{min}$, $I(B_z)$) is 99.7\%, while the ss of CC(SYM-H$_{min}$, $I(E_y)$) is 99.9\%.

\begin{figure}
  \centering
  \includegraphics[width=0.8\textwidth]{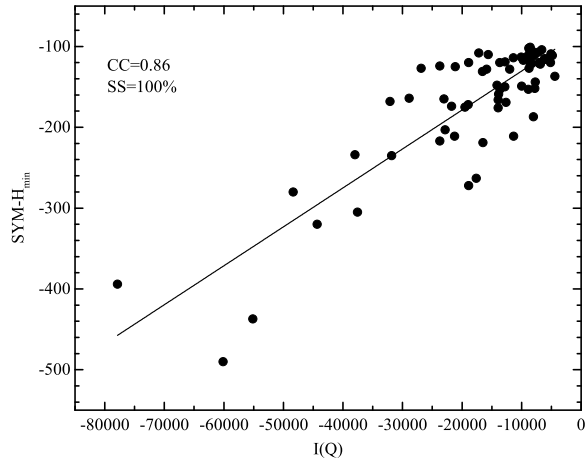}
  \caption{The CC between SYM-H and time integral of Q during storm main phases}\label{fig-02}
\end{figure}

The CC between SYM-H$_{min}$ and $I(Q)$ for major storms was calculated and the result is shown in Figure \ref{fig-02}. As shown in Figure\ref{fig-02}, CC(SYM-H$_{min}$, $I(Q)$) is 0.86. The derived ss of CC(SYM-H$_{min}$, $I(Q)$) is 100\%. It is evident that CC(SYM-H$_{min}$, $I(Q)$) is much larger than CC(SYM-H$_{min}$, $I(E_y)$), suggesting that Q is much more important than $E_y$. According to formula (\ref{eq-05}), we can easily judge that solar wind dynamic pressure is an important parameter affecting the intensity of major geomagnetic storms.

\section{Discussion and Summary} 
      \label{S-Summary}

\begin{figure}
  \centering
  \includegraphics[width=0.8\textwidth]{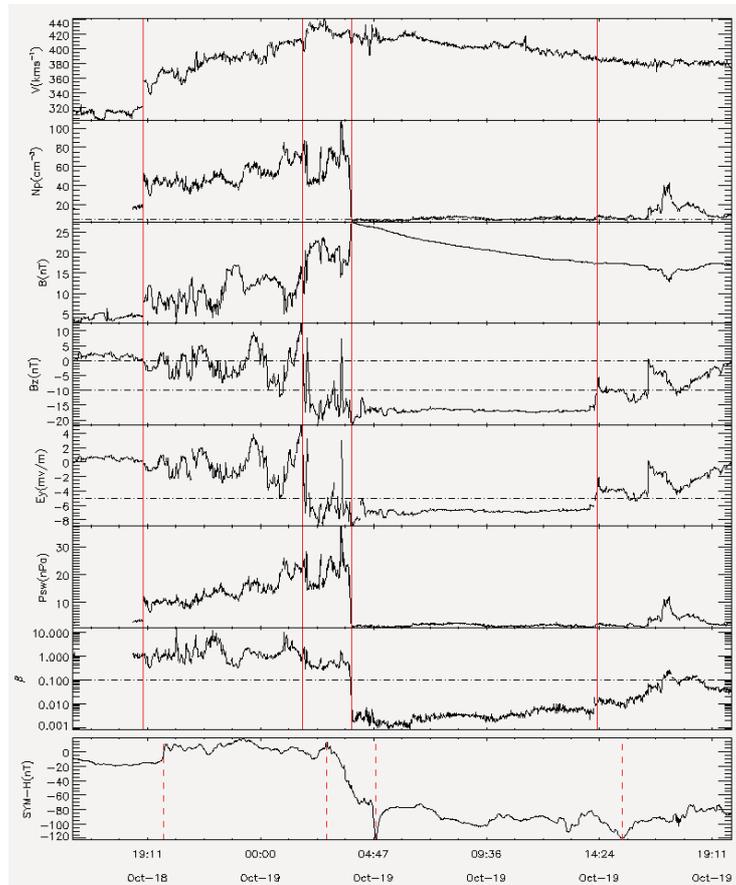}
  \caption{The solar wind parameters observed by ACE spacecraft during 18-19, October 1998. Form top to bottom, it shows solar wind speed, density, magnetic field strength, z-component of the magnetic field, solar wind electric field, solar wind dynamic pressure, proton $\beta$ and SYM-H index respectively. }\label{fig-03}
\end{figure}

In the current literature, it is usually accepted that large and long duration $B_s$ ($B_s>10 nT$ for more than 3 h) will lead to a major geomagnetic storm (Gonzalez and Tsurutani, 1987). However, the statistical results of the present study suggest that SYM-H$_{min}$ has a poor correlation with $I(B_z)$, indicating that strong and long lasting $B_s$ can not independently trigger a major storm. How should one understand this kind of result? An example used to explain this kind of result is shown in Figure \ref{fig-03}. A major geomagnetic storm that occurred on 18 October 1998 was used to explain this phenomenon. The solar source of a major geomagnetic storm is a coronal mass ejection (CME) with a source location at N10E10; the CME erupted at 10:04 UT on 15 October 1998, with a projected speed of 362 km/s (Zhang et al 2007). When the CME reached the interplanetary space around the Earth, a shock and sheath region as well as a magnetic cloud (MC) were observed by spaceship ACE. Solar wind parameters responsible for the major geomagnetic storm are shown in Figure 3. An interplanetary shock indicated by the first vertical red solid line in Figure \ref{fig-03} was observed by ACE spacecraft at 19:00 UT, 18 October 1998. When the shock reached the magnetosphere at 19:52 UT, it caused a storm sudden commencement, which was indicated by the first vertical red dashed line in Figure \ref{fig-03}. The solar wind between the second and third vertical solid lines is a sheath, while the solar wind between the third and fourth vertical solid red lines is an MC, which is the interplanetary structure associated with enhanced magnetic field strength. Long and smooth rotation of the magnetic field vector and low proton temperature (Burlaga et al. 1981). We can see from Figure 3 that the SYM-H index decreased quickly due to the sheath. The average Bs and Ey of the MC between the third and fourth vertical solid lines are 16.8 nT and 6.8 mV/m, respectively, and the time duration between the third and fourth vertical solid lines is longer than 10 h. According to the criteria proposed by Gonzalez and Tsurutani (1987), the MC should trigger intense geomagnetic storms. However, as shown in Figure 3, the MC has a weak geoeffectiveness. The averaged solar wind dynamic pressure of the MC is 1.43 nPa. This may be the reason why the MC only has a weak geoeffectiveness. The averaged solar wind dynamic pressure in the sheath is 18.5 nP, which is much larger than that of the MC. The averaged $B_s$ and $E_y$ of the sheath are 13.8 nT and 5.9 mV/m respectively, and the duration of the sheath is slightly more than 2 h. It is evident that the time duration of the MC is longer than that of the sheath. The average $B_s$ and $E_y$ of the sheath are smaller than those of the MC. However, the geoeffectiveness of the sheath is much larger than that of the MC. It is evident that the major geomagnetic storm was caused by the sheath. Sheath plasma causing magnetic storms was first discussed by Tsurutani et al. (1988). This case supports the concept that solar wind dynamic pressure is an important parameter driving the intensity of associated geomagnetic storms. Figure 3 shows that a large and long duration southward interplanetary magnetic field or solar wind electric field may not trigger a major geomagnetic storm if the dynamic pressure is very low.

According to the formula (\ref{eq-05}), if solar wind dynamic pressure is equal to 3 nPa during the main phases of geomagnetic storms, then CC(SYM-H$_{min}$, $I(E_y)$) will be equal to CC(SYM-H$_{min}$, $I(Q)$). We find that solar wind dynamic pressures during the main phases of all major geomagnetic storms studied in this paper are larger than 3 nPa. This should be the reason why CC(SYM-H$_{min}$, $I(Q)$) is larger than CC(SYM-H$_{min}$, $I(E_y)$). The statistical results of the present study also provide evidence that the injection term of the ring current proposed by Wang et al. (2003) is more reasonable than that proposed by Burton et al.(1975), namely that the injection term of the ring current should be a function of southward IMF, solar wind speed and solar wind density, rather than only a linear function of solar wind electric field. Various coupling functions were proposed by Gonzales et al. (1989). We note that there might be a coupling function better than the injection term Q reported by Wang et al. (2003). However, identifying this coupling function is beyond the scope of the present study.

Solar wind dynamic pressure is $n_pV_{sw}^2$, where $n_p$ is solar wind density. Solar wind dynamic pressure larger than 3 nPa demands that solar wind density should satisfy the condition $n_p>3/V_{sw}^2$. Hence, solar wind density is dependent on solar wind speed, because lower solar wind speed requires higher solar wind density so that solar wind dynamic pressure can be larger than 3 nPa and vice versa.

When solar wind reaches the magnetosphere, three solar wind parameters, $B_z$, $V_{sw}$ and $n_p$, will interact with the magnetosphere simultaneously. The CC between a single solar wind parameter and the intensity of the associated geomagnetic storm, or the CC between the combination of two solar wind parameters and the intensity of the associated geomagnetic storm, has no physical significance because any solar wind parameter cannot be removed from the interaction between solar wind and the magnetosphere. It is evident that solar wind density cannot independently drive the ring current (O'Brien and McPherron 2000a), nor the southward interplanetary magnetic field, solar wind speed, or solar wind electric field.

The correlation coefficient between singe solar wind parameter and the intensity of associated geomagnetic storm, or the correlation coefficient between the combination of two solar wind parameters and the intensity of associated geomagnetic storm, has no physical significance because any solar wind parameter can not be removed from the interaction between solar wind and the magnetosphere. It is evident that solar wind density can not independently drive the ring current (O'Brien and McPherron 2000a), nor the southward interplanetary magnetic filed, solar wind speed and even solar wind electric field.

The results from the present study are summarized as:

$\int_{t_s }^{t_e}B_z\mathrm{d}t$, $\int_{t_s }^{t_e}E_y\mathrm{d}t$ and $\int_{t_s }^{t_e}Q\mathrm{d}t$ make a small, moderate and crucial contributions to the intensity of associated major geomagnetic storm, respectively. The results provide statistical evidence that $E_y$ is more important for the intensity of associated geomagnetic storm than $B_z$, while Q is more important for the intensity of associated geomagnetic storm than $E_y$. The statistical results suggest that southward IMF, solar wind speed and solar wind density are all important for the occurrence of a major geomagnetic storm. Solar wind that has a strong geoeffectiveness requies that solar wind dynamic pressure is at least larger than 3 nPa, or demands that solar wind density should be at least larger than $>3$nPa$/V_{sw}^2$. Large and long duration $B_s$ alone cannot guarantee a major geomagnetic storm if solar wind dynamic pressure is very low, as large and long duration $B_s$ is not a full condition, only a necessary condition to trigger a major geomagnetic storm.

\begin{acks}
 We are very grateful to the anonymous referee for her/his review of our manuscript and for the helpful suggestions. We thank the ACE SWEPAM instrument team and the ACE Science Center for providing the ACE data. We thank Center for Geomagnetism and Space Magnetism, Kyoto University, for providing SYM-H index. This work is supported by the National Natural Science Foundation of China (Grant No. 41074132, 41274193, 41674166)
\end{acks}


\end{article}

\end{document}